\documentclass[aps,pra,floatfix,twocolumn]{revtex4}
\usepackage[dvips]{graphicx}

\usepackage{longtable}
\usepackage{dcolumn}
\usepackage[dvips]{graphicx}
\usepackage{bm}
\usepackage{bbm}

\usepackage{times}
\usepackage{nicefrac}
\usepackage{amsmath}
\usepackage{amsfonts}
\usepackage{amssymb}
\usepackage{amsthm}

\newcolumntype{.}{D{x}{}{-1}}

\usepackage{color}

\newcommand{\Za}{Z\alpha}

\begin{document}

\title{Theory of the Lamb shift in hydrogen and light hydrogen-like ions}

\author{Vladimir A. Yerokhin} \affiliation{Center for Advanced Studies,
        Peter the Great St.~Petersburg Polytechnic University, Polytekhnicheskaya 29,
        St.~Petersburg 195251, Russia}

\author{Krzysztof Pachucki}
\affiliation{Faculty of Physics, University of Warsaw,
             Pasteura 5, 02-093 Warsaw, Poland}

\author{Vojt\v{e}ch Patk\'o\v{s}}
\affiliation{Faculty of Mathematics and Physics, Charles University,  Ke Karlovu 3, 121 16 Prague
2, Czech Republic}

\begin{abstract}

Theoretical calculations of the Lamb shift provide the basis required for the determination of
the Rydberg constant from spectroscopic measurements in hydrogen. The recent high-precision
determination of the proton charge radius drastically reduced the uncertainty in the hydrogen
Lamb shift originating from the proton size. As a result, the dominant theoretical uncertainty
now comes from the two- and three-loop QED effects, which calls for further advances in their
calculations. We review the present status of theoretical calculations of the Lamb shift in
hydrogen and light hydrogen-like ions with the nuclear charge number up to $Z = 5$. Theoretical
errors due to various effects are critically examined and estimated.

\end{abstract}

\maketitle

\section{Introduction}

Hydrogen atom plays a special role in modern physics. As the simplest atomic system, hydrogen is
often considered to be an ideal testing ground for exploring limits of the theory based on
predictions of the bound-state quantum electrodynamics (QED). One of the important tests of theory
is the comparison of the proton charge radius values obtained from the Lamb shift in electronic and
muonic hydrogen. The $4.5\,\sigma$ discrepancy between these values, known as the proton radius
puzzle \cite{pohl:10,antognini:13}, attracted large attention of the scientific community. This
discrepancy could indicate violation of the lepton universality and existence of interactions not
accounted for in the Standard Model. Such a possibility is still open, although recent experiments
on electronic hydrogen \cite{beyer:17,fleurbaey:18,pohl:talk,hessels:talk} hint at existence of
unknown systematic effects in hydrogen measurements rather than at new physics.

Another important role of hydrogen is that comparison of theory and experiment for its transition
energies is used \cite{mohr:16:codata} for determining the Rydberg constant, which is one of the
most accurately known fundamental constants today. If one adopts the proton charge radius
determined from the muonic hydrogen \cite{antognini:13}, the uncertainty of the Rydberg constant is
defined by the currently available theory of the hydrogen Lamb shift.

Precise spectroscopy of light hydrogen-like ions may also provide determinations of the Rydberg
constant in the foreseeable future. Such determinations will be independent on the proton radius
and systematic effects in the hydrogen spectroscopy. Helium isotopes look most promising in this
respect, because of high-precision results for nuclear radii expected soon from experiments on
muonic helium \cite{schmidt:18}. We mention here the ongoing projects of measuring the $1S$--$2S$
transition energy in He$^+$ pursued in Garching \cite{udem:priv} and in Amsterdam
\cite{eikema:priv}, which require improved theoretical predictions for the helium Lamb shift.

Motivated by the needs outlined above, in the present work we summarize the presently available
theory for the Lamb shift of hydrogen and light hydrogen-like ions with the nuclear charge up to $Z
= 5$. This summary is intended as an update of the CODATA review of the hydrogen theory
\cite{mohr:16:codata}. In particular, we perform a reanalysis of results available for the
higher-order two-loop QED corrections, which presently define the theoretical uncertainty of the
Lamb shift. Results for the nuclear recoil effect are significantly improved by taking into account
recent nonperturbative calculations \cite{yerokhin:15:recprl,yerokhin:16:recoil}. The nuclear
finite size and nuclear polarizability effects are reformulated, according to recent theoretical
developments \cite{pachucki:18,tomalak:18}.

Relativistic units $m = \hbar = c = 1$ are used throughout this paper (where $m$ is the electron
mass). In these units the electron rest mass energy $mc^2 = 1$, so that all energy corrections
appear to be dimensionless. In order to convert any energy correction in relativistic units to
arbitrary units, it is sufficient to multiply it by $2{\cal R}/\alpha^2$, where ${\cal R} =
hcR_{\infty}$ is the Rydberg energy and $R_{\infty}$ is the Rydberg constant. While $m = 1$ in our
units, we will write $m$ explicitly when it enters dimensionless ratios, such as $m/M$ and $m_r/m$.

\section{Binding energy}

We consider the binding energy $E_{njl}$ of an electronic state with quantum numbers $n$, $j$, and
$l$ in a light hydrogen-like atom. If the atomic nucleus has a nonzero spin $I$, the energy level
$|njl\big>$ is splitted by the interaction with the nuclear magnetic moment according to values of
the total angular momentum $F$, $|njlF\big>$. In this case, we define the binding energy $E_{njl}$
as a centroid averaged over all hyperfine-structure (hfs) components,
\begin{align} \label{eq0a}
E_{njl} = \frac{\sum_{F}(2F+1)\, E_{njlF}}{\sum_{F} (2F+1)}\,.
\end{align}
The interaction with the dipole nuclear magnetic moment, responsible for the hyperfine structure,
does not contribute to $E_{njl}$ in the first order. There is, however, a second-order hfs effect
that shifts (slightly) the centroid energy $E_{njl}$. It manifests itself as a nuclear-spin
dependent recoil correction and is addressed in Sec.~\ref{sec:recoil}.

The goal of the present paper is to summarize the presently available theory for the binding energy
$E_{njl}$ of the $1S$, $2S$, and $2P_{1/2}$ states of light hydrogen-like atoms. The hyperfine
splitting of energy levels will not be discussed. For the $nS$ states it was investigated in detail
in Ref.~\cite{jentschura:06:hfs}; a review of the hfs of the higher-$l$ states is available in
Ref.~\cite{horbatsch:16}.

The binding energy of a light hydrogen-like atom is usually represented as a sum of three
contributions,
\begin{align} \label{eq1a}
E_{njl} &\ =  E_{ D} + E_{ M} + E_{ L}\,,
\end{align}
where $E_{ D}$ is the Dirac point-nucleus biding energy in the nonrecoil limit, $E_{ M}$ is the
correction containing the dominant part of the nuclear recoil effect, and $E_{ L}$ is the Lamb
shift. We note that the total recoil effect is thus distributed between $E_M$ and $E_L$ ($E_M$
being the dominant part and smaller corrections being ascribed to the Lamb shift $E_L$). This
distribution is not unique and done differently in the literature.

The Dirac point-nucleus nonrecoil binding energy $E_{ D}$ is given by
\begin{align} \label{eq2}
E_{ D} =
\sqrt{1 - \frac{(\Za)^2}{N^2}} - 1
\,,
\end{align}
where
\begin{align} \label{eq:N}
N &\ = \sqrt{(n_r+\gamma)^2 + (\Za)^2}\,,\
\end{align}
$\gamma = \sqrt{\kappa^2-(\Za)^2}$, $n_r = n - |\kappa|$ is the radial quantum number, $n$ is the
principal quantum number, and $\kappa = (l-j)(2j+1)$ is the angular momentum-parity quantum number.

The leading recoil correction $E_{ M}$ is
\begin{align} \label{eq3}
E_{ M} = \frac{m}{M}\frac{(\Za)^2}{2\,N^2} - \left( \frac{m}{M}\right)^2 \,
   \frac{(\Za)^2}{2\,n^2}\,\frac{m_r}{m} \,,
\end{align}
where $M$ is the nuclear mass and $m_r = mM/(m+M)$ is the reduced mass. All further recoil
corrections are ascribed to the Lamb shift $E_L$. The first part of $E_M$ comprises the complete
$m/M$ recoil effect to orders $(\Za)^2$ and $(\Za)^4$ and, in addition, corrections of order
$(\Za)^6$ and higher that can be obtained from the Breit Hamiltonian. The second part of $E_M$ is
the nonrelativistic recoil correction of second and higher orders in $m/M$. In the nonrelativistic
limit, the sum $E_D+E_M$ reduces to the Schr\"odinger energy eigenvalue,
\begin{align}
E_D+E_M = \frac{m_r}{m} \frac{(\Za)^2}{2\,n^2} + \ldots\,,
\end{align}
where $\ldots$ represents contributions of order $(\Za)^4$ and higher.

Our choice of $E_{M}$ (and, therefore, our definition of the Lamb shift $E_L$) follows
Ref.~\cite{yerokhin:15:Hlike} and differs slightly from the popular definition
\cite{sapirstein:90:kin} based on the Barker-Glover formula \cite{barker:55} and, as a consequence,
from the definition of the CODATA review \cite{mohr:16:codata} ({\em cf.} Eqs.~(25) and (26)
therein). The reason for this difference was the need for a simple and concise definition valid for
an arbitrary nucleus, whereas the Barker-Glover formula is valid only for the spin-$1/2$ nucleus.
Both definitions are equivalent through orders $(m/M)(\Za)^{2+n}$ and $(m/M)^{2+n}(\Za)^2$, with $n
\ge 0$. The difference is that our present definition of Eq.~$(\ref{eq3})$ does not contain any
contribution of order $(m/M)^2(\Za)^4$ (which depends on the nuclear spin) or any spurious
higher-order terms. The correction of order $(m/M)^2(\Za)^4$ is included into the Lamb shift; it is
given by the first line of Eq.~(\ref{rec:3}).

Another difference in definitions in the literature is associated with the off-diagonal hfs
correction, which is small but relevant on the level of the experimental interest for the $l>0$
states \cite{brodsky:67}. In the old Lamb-shift measurements (in particular,
Ref.~\cite{lundeen:81}), this correction was subtracted from the experimental result. Reviews
\cite{mohr:16:codata,yerokhin:15:Hlike} do not discuss it, thus excluding it from the definition of
the Lamb shift. The review \cite{horbatsch:16} includes this correction [see Eq.~(30) therein] but
ascribes it to the hyperfine splitting. A part of the off-diagonal hfs correction shifts the
centroid energy $E_{njl}$ and thus needs to be included into the definition of the Lamb shift. The
corresponding contribution is given by Eq.~(\ref{rec:4}).

We now turn to examining various effects that contribute to the Lamb shift $E_L$.

\section{QED effects}

\subsection{One-loop QED effects}

The one-loop QED effects for the point nuclear charge are represented as
\begin{align} \label{se:1}
E_{\rm QED1 } = &\ \frac{\alpha}{\pi}\, \frac{(\Za)^4}{n^3}\,\left(\frac{m_r}{m} \right)^3\,
  \nonumber \\ &
   \times
  \Bigl[ F_{\rm SE }(\Za) + F_{\rm VP }(\Za)\Bigr]\,,
\end{align}
where the functions $F_{\rm SE }(\Za)$ and $F_{\rm VP }(\Za)$ correspond to the one-loop
self-energy and vacuum-polarization, respectively.

The $\Za$ expansion of the electron self-energy is given by
\begin{align} \label{eq:se:za}
F_{\rm SE }(\Za) &\ = L\,A_{41} + A_{40} + (\Za)\,A_{50}
  \nonumber \\ &
+ (\Za)^2\,\biggl[L^2\,A_{62} + L\,A_{61}+ G_{\rm SE,pnt}(\Za)\biggr]\,,
\end{align}
where $L = \ln\left[(m/m_r)(\Za)^{-2}\right]$ and  $G_{\rm SE }(\Za) = A_{60} + \ldots$ is the
remainder that contains all higher-order expansion terms in $\Za$. The coefficients of the $\Za$
expansion in Eq.~(\ref{eq:se:za}) are well known. They are discussed, e.g., in
review~\cite{eides:01} and summarized in Table~\ref{tab:se}. Numerical results for the remainder
function are obtained by Jentschura and Mohr \cite{jentschura:99:prl,jentschura:01:pra} and listed
in Table~\ref{tab:qed1:ho}. Results for $Z=0$ correspond to the coefficient $A_{60}$; they were
taken from Ref.~\cite{jentschura:05:sese}.

The $\Za$ expansion of the vacuum-polarization correction is given by
\begin{align} \label{eq:vp:za}
F_{\rm VP }(\Za)&\  = -\frac{4}{15}\,\delta_{\ell0} + \frac{5}{48}\pi(\Za)\,\delta_{\ell0}
+ (\Za)^2 \,
  \nonumber \\ &
   \times
    \biggl[-\frac{2}{15} L\, \delta_{\ell0}
+ G_{\rm Ueh}(\Za)
+G_{\rm WK}(\Za)\biggr]\,,
\end{align}
where $G_{\rm Ueh}(\Za)$ and $G_{\rm WK}(\Za)$ are the higher-order remainder functions induced by
the Uehling and Wichmann-Kroll parts of the vacuum polarization, respectively. Numerical results
for the remainder functions are listed in Table~\ref{tab:qed1:ho}. The Wichmann-Kroll part of the
vacuum polarization was calculated with help of the approximate potential based on the analytical
expansions of Whittaker functions from Ref.~\cite{manakov:12:vgu}. The uncertainty due to
approximations in the potential is negligible at the level of current interest. In the limit $Z\to
0$, results for the higher-order remainders are (see review \cite{eides:01} for details)
\begin{align}
&G_{\rm Ueh}(Z = 0, 1S) = \frac{4}{15} \Big(\ln 2 - \frac{1289}{420}\Big)\,,\\
&G_{\rm Ueh}(Z = 0, 2S) = -\frac{743}{900} \,,\\
&G_{\rm Ueh}(Z = 0, 2P_{1/2}) = -\frac{9}{140}\,,\\
&G_{\rm WK}(Z = 0) = \Big(\frac{19}{45} - \frac{\pi^2}{27}\Big)\,\delta_{\ell0}\,.
\end{align}

The vacuum-polarization induced by the $\mu^+\mu^-$ virtual pairs is given by
\cite{eides:95:pra,karshenboim:95}
\begin{align}\label{qed3:2}
E_{\mu \rm VP} = \left(\frac{m}{m_{\mu}}\right)^2 \frac{\alpha}{\pi}\,\frac{(\Za)^4}{n^3}\,
\left(\frac{m_r}{m} \right)^3\,
\left( -\frac{4}{15}\right)\,\delta_{\ell0}\,,
\end{align}
where $m_{\mu}$ is the muon mass.

The hadronic vacuum-polarization correction is of the same order as the muonic vacuum polarization
and is given by \cite{friar:99}
\begin{align}\label{qed3:3}
E_{ \rm hadVP} = 0.671\,(15)\ E_{\mu \rm VP}\,.
\end{align}

\begin{table*}
\caption{Coefficients of the $Z\alpha$ expansion of the one-loop electron self-energy in Eq.~(\ref{eq:se:za}). \label{tab:se}}
\begin{ruledtabular}
\begin{tabular}{lccccc}
\multicolumn{1}{c}{Term} &  \multicolumn{1}{c}{} & \multicolumn{1}{c}{$1S$} & \multicolumn{1}{c}{$2S$}& \multicolumn{1}{c}{$2P_{1/2}$}\\
\hline\\[-5pt]
$A_{41}$  &    $\frac43\,\delta_{\ell0}$ & $\frac43$ & $\frac43$ & $0$ \\
$A_{40}$  & $-\frac43\,\ln k_0(n,l) + \frac{10}{9}\,\delta_{\ell0}
    - \frac{m/m_r}{2\kappa(2l+1)}
(1-\delta_{\ell0})$
    & $-2.867\,726\,964$ & $-2.637\,915\,413$   &  $-0.126\,644\,388\, (m/m_r)$\\
$A_{50}$  & $\left( \frac{139}{32} - 2 \ln 2\right)\,\pi \,\delta_{\ell0}$
    & $\ \ \ 9.291\,120\,908$ & $\ \ \ 9.291\,120\,908$ & $0$ \\
$A_{62}$  & $-\delta_{\ell0}$ & $-1$ & $-1$ & $0$\\
$A_{61}$  & $ 4 \Bigl(\frac{4}{3}\ln 2 + \ln \frac{2}{n} +\psi(n+1)-\psi(1) - \frac{601}{720} -
\frac{77}{180n^2}\Bigr)\delta_{\ell0}$
  & \ \ \ 5.419\,373\,685 & \ \ \ 5.930\,118\,296 & $ \ \ \ 0.572\,222\,222$ \\
          & $+\biggl[
           \frac{n^2-1}{n^2}\bigl( \frac{2}{15}+\frac13 \delta_{j,1/2}\Bigr)
           + 8 \frac{3 - l (l+1)/n^2}{3(2l+3)l(l+1)(4l^2-1)}\biggr] (1-\delta_{\ell0})$ \\
\end{tabular}
\end{ruledtabular}
\end{table*}
\begin{table*}
\caption{Results for the higher-order remainder functions
$G_{\rm SE}$, $G_{\rm Ueh}$, and $G_{\rm WK}$ in Eqs.~(\ref{eq:se:za}) and (\ref{eq:vp:za}).
\label{tab:qed1:ho}}
\begin{ruledtabular}
\begin{tabular}{lddd}
\multicolumn{1}{c}{$Z$} & \multicolumn{1}{c}{$1S$} & \multicolumn{1}{c}{$2S$}& \multicolumn{1}{c}{$2P_{1/2}$}\\
\hline\\[-5pt]
\multicolumn{4}{l}{Self-energy:}\\
 0   & -30.924\,149\,46\,(1)&-31.840\,465\,09\,(1)& -0.998\,904\,40 \\
 1   & -30.290\,24\,(2)  &  -31.185\,15\,(9)    &  -0.973\,45\,(19) \\
 2   & -29.770\,967\,(5) &  -30.644\,66\,(5)    &  -0.949\,40\,(5)  \\
 3   & -29.299\,170\,(2) &  -30.151\,93\,(2)    &  -0.926\,37\,(2)  \\
 4   & -28.859\,222\,(1) &  -29.691\,27\,(1)    &  -0.904\,12\,(1)  \\
 5   & -28.443\,372\,(1) &  -29.255\,033\,(8)   &  -0.882\,478\,(8)  \\[3pt]
\multicolumn{4}{l}{Vacuum-polarization, Uehling:}\\
 0  &  -0.633\,573  &   -0.825\,556  &      -0.064\,286 \\
 1 	&  -0.618\,724  &  	-0.808\,872  &  	-0.064\,006  \\
 2 	&  -0.607\,668  &  	-0.796\,118  &  	-0.063\,768  \\
 3 	&  -0.598\,207  &  	-0.785\,075  &  	-0.063\,567  \\
 4 	&  -0.589\,838  &  	-0.775\,230  &  	-0.063\,399  \\
 5 	&  -0.582\,309  &  	-0.766\,322  &  	-0.063\,262  \\[3pt]
\multicolumn{4}{l}{Vacuum-polarization, Wichmann-Kroll:}\\
 0  &    0.056\,681  &    0.056\,681  &      0 \\
 1  &  	 0.055\,721  &    0.055\,721  &  	 0.000\,002  \\
 2  &  	 0.054\,823  &    0.054\,824  &  	 0.000\,006  \\
 3  &  	 0.053\,978  &    0.053\,983  &  	 0.000\,012  \\
 4  &  	 0.053\,178  &    0.053\,188  &  	 0.000\,020  \\
 5  &  	 0.052\,418  &    0.052\,437  &  	 0.000\,030  \\
\end{tabular}
\end{ruledtabular}
\end{table*}

%%%%%%%%%%%%%%%%%%%%%%%%%%%%%%%%%%%%%%%%%%%%%%
\subsection{Two-loop QED effects}

The two-loop QED correction is expressed as
\begin{align}\label{qed2:1}
E_{\rm QED2} = \left(\frac{\alpha}{\pi}\right)^2\, \frac{(\Za)^4}{n^3}\, \left( \frac{m_r}{m}\right)^3\, F_{\rm QED2}(\Za)\,,
\end{align}
where the function $F_{\rm QED2}$ is given by
\begin{align} \label{eq:twoloop:za}
F_{\rm QED2}(\Za) =& \  B_{40} + (\Za)\,B_{50} + (\Za)^2\bigl[ B_{63}\, L^3
  \nonumber \\ &
+ B_{62}\, L^2 + B_{61}\, L + G_{\rm QED2}(\Za)\bigr]\,,
\end{align}
and $G_{\rm QED2}(\Za) = B_{60} + \ldots$ is the remainder that contains all higher-order expansion
terms in $\Za$.

The two-loop QED correction is conveniently divided into three parts: the two-loop self-energy
(SESE), the two-loop vacuum-polarization (VPVP), and the mixed self-energy and vacuum-polarization
(SEVP),
\begin{align} \label{qed2:2}
F_{\rm QED2} = F_{\rm SESE} + F_{\rm SEVP} + F_{\rm VPVP}\,.
\end{align}
Coefficients of the $\Za$ expansion of the individual two-loop corrections for the states under
consideration are summarized in Table~\ref{tab:twoloop}, for details see recent studies
\cite{pachucki:01:pra,pachucki:03:prl,czarnecki:05:prl,jentschura:05:sese,dowling:10,czarnecki:16}
and references to earlier works therein. We note that the analytical result for the $B_{61}$
coefficient derived in Ref.~\cite{pachucki:01:pra} was incomplete; one missing piece was added
later in Ref.~\cite{jentschura:05:sese} and another, in Ref.~\cite{czarnecki:16}. The listed value
of $B_{61}$ for the $S$ states differs from that given in
Refs.~\cite{mohr:16:codata,yerokhin:15:Hlike} by $-43/36 + 133 \pi^2/864 = -0.134567\ldots$, which
is the light-by-light contribution from Ref.~\cite{czarnecki:16}. Numerical values for the
delta-function correction to the Bethe logarithm ${\cal N}(nS)$ and ${\cal N}(nP)$ that enter
$B_{61}$ can be found in Refs.~\cite{jentschura:03:jpa,jentschura:05:sese}.

The two-loop higher-order remainder $G_{\rm QED2}$ is only partly known up to now. Its $\Za$
expansion has the form
\begin{align}  \label{qed2:3}
G_{\rm QED2}(\Za) = B_{60} + (\Za) \big[ B_{72}\, L^2 + B_{71}\, L + \ldots \big]\,.
\end{align}

The dominant part of the coefficient $B_{60}$ comes from the two-loop self-energy. It was
calculated for the $1S$ and $2S$ states by Pachucki and Jentschura \cite{pachucki:03:prl}, with the
result
\begin{align} \label{b60}
B_{60}(1S,{\rm SESE}) = -61.6\,(9.2)\,,\\
B_{60}(2S,{\rm SESE}) = -53.2\,(8.0)\,,
\end{align}
where the uncertainty comes from omitted contributions. The complete $n$ dependence of $B_{60}(nS)$
was calculated in Refs.~\cite{czarnecki:05:prl,jentschura:05:sese}.  For the $nP$ states, the
coefficient $B_{60}$ was calculated in Ref.~\cite{jentschura:06:sese}. The results for the SESE and
SEVP corrections and the $2P_{1/2}$ state are
\begin{align} \label{b60p}
B_{60}(2P_{1/2},{\rm SESE}) &\ = -1.5\,(3)\,, \\
B_{60}(2P_{1/2},{\rm SEVP}) &\ = -0.016\,571\ldots \,.
\end{align}
We use opportunity to correct a mistake in Ref.~\cite{jentschura:06:sese} for the VPVP correction
(given by Eqs.~(A3) and (A6) of that work). The corrected results are
\begin{align} \label{b60p2}
B_{60}(nP_{1/2},{\rm VPVP}) &\ = -\frac{713}{2025}\left(1-\frac1{n^2}\right)\,,\\
B_{60}(nP_{3/2},{\rm VPVP}) &\ = -\frac{401}{4050}\left(1-\frac1{n^2}\right)\,.
\end{align}

The logarithmic coefficients $B_{72}$ and $B_{71}$ in Eq.~(\ref{qed2:3}) were recently investigated
in Ref.~\cite{karshenboim:18}. The leading logarithmic coefficient $B_{72}$ was derived as
\begin{align} \label{b72}
&B_{72}({\rm SESE}) = \Big( -\frac{139}{48} + \frac43 \ln 2\Big)\,\pi\,\delta_{\ell0}\,,\\
&B_{72}({\rm SEVP}) = -\frac{5}{72}\,\,\pi\,\delta_{\ell0}\,,\\
&B_{72}({\rm VPVP}) = 0\,.
\end{align}
The next coefficient $B_{71}$ was obtained for the $nP$ states, with the result
\begin{align} \label{b71}
&B_{71}({\rm SESE},nP) = \Big( \frac{139}{144} - \frac49 \ln 2\Big)\,\pi\,\frac{n^2-1}{n^2}\,\,,\\
&B_{71}({\rm SEVP},nP) = \frac{5}{216}\,\pi\,\frac{n^2-1}{n^2}\,\,,\\
&B_{71}({\rm VPVP},nP) = 0\,.
\end{align}
Ref.~\cite{karshenboim:18} also reported the $n$ dependence of $B_{71}(nS)$.

Calculations of the SESE part of the higher-order remainder, $G_{\rm SESE}$, were carried out to
all orders in $\Za$ for hydrogen-like ions with $Z\ge 10$ \cite{yerokhin:03:prl,yerokhin:03:epjd}.
The latest results were obtained in Ref.~\cite{yerokhin:09:sese} for $Z < 30$ and in
Ref.~\cite{yerokhin:18:sese} for $Z \ge 30$. The extrapolation of the all-order $1S$ results down
to $Z = 1$ reported in Ref.~\cite{yerokhin:09:sese} showed only a marginal agreement with the
analytical value (\ref{b60}). A possible reason for this could be a large contribution from the
unknown logarithmic coefficient $B_{71}$.

In the present work, we merge together the numerical and analytical results, in order to obtain the
presumably best values for the higher-order remainder. Specifically, for the $1S$ state, we fit the
numerical all-order data for $Z\ge 15$ from Refs.~\cite{yerokhin:09:sese,yerokhin:18:sese} to the
form
\begin{align}
G_{\rm SESE}(1S) = &\ B_{60} + B_{72}(\Za)\ln^2(\Za)^{-2}
  \nonumber \\ &
+ b_{71}(\Za)\ln(\Za)^{-2}
 + (\Za)\,{\rm pol}(\Za)\,,
\end{align}
where $B_{60}$ and $B_{72}$ are given by Eqs.~(\ref{b60}) and (\ref{b72}), and ${\rm pol}(\Za)$
denotes a polynomial in $\Za$. $b_{71}$ and the coefficients of the polynomial are fitting
parameters. The uncertainty was obtained by varying ({\em i}) $B_{60}$ within its error bars
(\ref{b60}), ({\em ii}) numerical data within their error bars, and ({\em iii}) the length of the
polynomial and the number of data points included. The higher-order remainder for the $2S$ state
was obtained by adding to $G_{\rm SESE}(1S)$ the difference $G_{\rm SESE}(2S)-G_{\rm SESE}(1S)$, as
fitted in Ref.~\cite{yerokhin:18:sese}. For the $2P_{1/2}$ state, we merged together the analytical
results (\ref{b60p}) and (\ref{b71}) and numerical data from Ref.~\cite{yerokhin:18:sese}. The
uncertainty was obtained by quadratically adding the error of the $B_{60}$ coefficient and one half
of the leading logarithmic $B_{71}$ contribution. The obtained results for the higher-order SESE
remainder are summarized in Table~\ref{tab:qed2:ho}.

Calculations of the SEVP and VPVP corrections were performed in Ref.~\cite{yerokhin:08:twoloop} to
all orders in $\Za$. Results for the higher-order remainder $G_{\rm SEVP}$ listed in
Table~\ref{tab:qed2:ho} were obtained from Tables I and IV of Ref.~\cite{yerokhin:08:twoloop},
after subtracting contributions of the leading $\Za$-expansion coefficients and keeping in mind
that the light-by-light (LBL) contribution was not included in numerical calculations and thus
should not be subtracted. The uncertainty of the SEVP contribution comes from the missing LBL
contribution. It was estimated for the $S$ states as one half of the LBL $B_{61}$ contribution,
calculated in Ref.~\cite{czarnecki:16}. For the $P$ states, we assume the uncertainty to be
negligible.

The results for the higher-order remainder $G_{\rm VPVP}$ listed in Table~\ref{tab:qed2:ho} were
obtained from Tables II and III of Ref.~\cite{yerokhin:08:twoloop}, after subtracting contributions
of the leading $\Za$-expansion coefficients summarized in Table~\ref{tab:twoloop}. The uncertainty
due to omitted higher-order K\"all\'en-Sabry contributions is assumed to be negligible at the level
of present interest.

For the $1S$ state of hydrogen, our result for the two-loop higher-order remainder is $G_{\rm QED2}
= -92(13)$, which is slightly lower than the value adopted by CODATA~2016 of $-81(20)$
\cite{mohr:16:codata}.

\begin{table*}
\caption{Coefficients of the $Z\alpha$ expansion of the two-loop QED effects in Eq.~(\ref{eq:twoloop:za}).
$\zeta(n)$ denotes the Riemann zeta function, $\psi(n)$ is the digamma function, $\gamma_E$ is Euler's constant,
${\cal N}(nL)$ is a delta-function correction to the Bethe logarithm, defined by Eq.~(4.21a) of Ref.~\cite{jentschura:05:sese}.
\label{tab:twoloop}}
\begin{ruledtabular}
\begin{tabular}{lccccc}
\multicolumn{1}{c}{Term} &  \multicolumn{1}{c}{} & \multicolumn{1}{c}{$1S$} & \multicolumn{1}{c}{$2S$}& \multicolumn{1}{c}{$2P_{1/2}$}\\
\hline\\[-5pt]
\multicolumn{1}{c}{\bf SESE}\\[5pt]
$B_{40}$  &  $\Big[-\frac{163}{72}-\frac{85}{216}\,\pi^2
   + \frac32 \, \pi^2 \ln 2\, -\frac94\, \zeta(3)  \Big]\,\delta_{\ell0}$
    & $1.409\,244$ & $1.409\,244$ & $0.114\,722\,(m/m_r)$ \\
          & $-  \Big[ -\frac{31}{16}  + \frac{5}{12}\,{\pi}^2 -
  \frac12 {\pi }^2\,\ln 2 + \frac34\,\zeta(3) \Big]
    \frac{m/m_r}{\kappa\,(2\,l+1)}\,\big(1-\delta_{l0}\big)$
     \\
$B_{50}$  &    unknown
    & $-24.265\,06\,(13)$  & $-24.265\,06\,(13)$ & 0 \\
$B_{63}$  & $-\frac{8}{27}\,\delta_{\ell0}$
    &  $-8/27$ &  $-8/27$ &  0 \\
$B_{62}$  &
$\frac{16}{9}\,\Big(
  \frac{13}{12}-\ln 2
     +\frac1{4n^2}-\frac1n-\ln n
      + \psi(n)+\gamma_E \Big)\,\delta_{\ell0}$
      & $-0.639\,669$ & $0.461\,403$ & $1/9$\\
          & $+ \frac{4}{27}\,\frac{n^2-1}{n^2}\,\delta_{\ell1}$\\
$B_{61}$  &
   $      \frac{4}{3}\,{\cal N}(nL) +
\bigg[\frac{15473}{2592} + \frac{1039}{432}\,{\pi }^2
  - \frac{152}{27}\,\ln 2 -
  \frac{2}{3}\,{\pi }^2\,\ln 2 + \frac{40}{9}\,\ln^2 2 +
   \zeta (3)
   $
   & $48.388\,913$ & $40.932\,915$ & $0.202\,220$ \\
   &
   $+\Big( \frac{80}{27}-\frac{32}{9}\,\ln 2 \Big)
   \Big( \frac34 +\frac1{4n^2}-\frac1n -\ln n +\psi(n)+\gamma_E
   \Big) \bigg]\,\delta_{\ell0}$
   \\
   &
   $ +\frac{n^2-1}{n^2}
    \Big(\frac{11}{81} + \frac13\delta_{j,1/2}-\frac{8}{27}\,\ln 2\Big)     \,\delta_{\ell1}
   $
   \\[5pt]
\multicolumn{1}{c}{\bf SEVP}\\[5pt]
$B_{40}$  &  $\left(-\frac{7}{81}+\frac{5\pi^2}{216}\right) \,\delta_{\ell0} $
    & $0.142\,043$ & $0.142\,043$ & $-0.005\,229\,(m/m_r)$\\
          &    $+ \left( \frac{119}{36}-\frac{\pi^2}{3}\right)\,\frac{j(j+1)-l(l+1)-3/4}{l(l+1)(2l+1)}
   \frac{m}{m_r}\,(1-\delta_{\ell0})$
    &  & & \\
$B_{50}$  &   unknown
    & $1.305\,370$ &$1.305\,370$ & 0 \\
$B_{63}$  & 0
    &  0 &  0 &  0 \\
$B_{62}$  &   $\frac{8}{45}\,\delta_{\ell0}$
    & $8/45$ & $8/45$ & 0 \\
$B_{61}$  &
$
\Big[-\frac{259}{1080}  + \frac{41\,{\pi }^2}{432} + \frac{16}{15}\ln 2 -\frac{32}{45}\,\Big(
\frac{3}{4}  + \frac{1}{4\,n^2} - \frac{1}{n} -
      \ln n + \psi(n)+ \gamma_E \Big)\Big] \,\delta_{\ell0}$
      & $1.436\,241$ & $0.995\,812$ &   $-0.044\,444$
      \\
      & $-\frac{2}{45}\,\delta_{\ell1}$\\[5pt]
\multicolumn{1}{c}{\bf VPVP}\\[5pt]
$B_{40}$  &  $-\frac{82}{81}\,\delta_{\ell0}$
    & $-82/81$ & $-82/81$ & $0$\\
$B_{50}$  & $\Big( \frac{7421-2625\pi}{6615} + \frac{52}{63}\ln2\Big)\,\pi\,\delta_{\ell0}$
   & $1.405241$ & $1.405241$ & 0 \\
$B_{63}$  & 0
    &  0 &  0 &  0 \\
$B_{62}$  & 0
    &  0 &  0 &  0 \\
$B_{61}$  & $-\frac{1097}{2025}  \,\delta_{\ell0}$
    &  $-0.541\,728$ &  $-0.541\,728$ &  0 \\
\end{tabular}
\end{ruledtabular}
\end{table*}

\begin{table}
\caption{Results for the two-loop higher-order remainder $G_{\rm QED2}$ in Eq.~(\ref{eq:twoloop:za}).
\label{tab:qed2:ho}}
\begin{ruledtabular}
\begin{tabular}{llll}
\multicolumn{1}{l}{$Z$} & \multicolumn{1}{c}{$1S$} & \multicolumn{1}{c}{$2S$}& \multicolumn{1}{c}{$2P_{1/2}$}\\
\hline\\[-5pt]
\multicolumn{1}{c}{\rm SESE}\\
  0    & $-61.6\,(9.2)$  & $-53.2\,(8.0)$  & $-1.5\,(3)$ \\
  1    & $-75.9\,(12.6)$ & $-61.2\,(12.6)$ & $-1.37\,(31)$ \\
  2    & $-82.6\,(9.9)$  & $-67.6\,(9.9)$  & $-1.28\,(31)$ \\
  3    & $-86.8\,(8.0)$  & $-71.7\,(8.0)$  & $-1.20\,(33)$ \\
  4    & $-89.7\,(6.7)$  & $-74.4\,(6.7)$  & $-1.13\,(34)$ \\
  5    & $-91.6\,(5.8)$  & $-76.3\,(5.8)$  & $-1.06\,(35)$ \\[5pt]
\multicolumn{1}{c}{\rm SEVP}\\
  0    &                &                 & $-0.01657$ \\
  1    & $-12.9\,(1.6)$ & $-11.3\,(1.6) $ & $-0.016\,(6)$ \\
  2    & $-11.8\,(1.4)$ & $-10.2\,(1.4) $ & $-0.015\,(5)$ \\
  3    & $-11.0\,(1.2)$ & $\ -9.4\,(1.2)$ & $-0.011\,(2)$ \\
  4    & $-10.5\,(1.2)$ & $\ -8.9\,(1.1)$ & $-0.007\,(2)$ \\
  5    & $-10.0\,(1.1)$ & $\ -8.4\,(1.1)$ & $-0.004\,(1)$ \\[5pt]
\multicolumn{1}{c}{\rm VPVP}\\
  0    &           &            & $-0.26407$ \\
  1    & $\ -2.76\,(2)$  &  $\ -3.37$ & $-0.263$ \\
  2    & $\ -2.70$ &  $\ -3.30$ & $-0.261$ \\
  3    & $\ -2.65$ &  $\ -3.24$ & $-0.260$ \\
  4    & $\ -2.61$ &  $\ -3.20$ & $-0.259$ \\
  5    & $\ -2.58$ &  $\ -3.16$ & $-0.258$ \\
\end{tabular}
\end{ruledtabular}
\end{table}

%%%%%%%%%%%%%%%%%%%%%%%%%%%%%%%%%%%%%%%%%%%%%%
\subsection{Higher-order QED effects}

The $\Za$ expansion of the three-loop QED correction is given by
\begin{align}
E_{\rm QED3} = &\ \left({\alpha\over\pi}\right)^3 {(Z\alpha)^4\over n^3}
\left(\frac{m_r}{m} \right)^3\,
 \Big[ C_{40}
 \nonumber \\ &
 + (\Za)\,C_{50}
 + (\Za)^2\,\Big( C_{62}\,L^2 + C_{61}\,L + \ldots\Bigr) \Big]\,,
\end{align}
The leading-order contribution $C_{40}$ was obtained in Refs.~\cite{laporta:96,melnikov:00} and is
given by
\begin{eqnarray}
C_{40} &=&
\bigg[ -{{568\,{\rm a_4}}\over{9}}+{{85\,\zeta(5)}\over{24}} \nonumber\\&&
-{{121\,\pi^{2}\,\zeta(3)}\over{72}} -{{84\,071\,\zeta(3)}\over{2304}} -{{71\,\ln ^{4}2}\over{27}}
\nonumber\\&& -{{239\,\pi^{2}\,\ln^{2}2}\over{135}} +{{4787\,\pi^{2}\,\ln 2}\over{108}}
+{{1591\,\pi^{4}}\over{3240}} \nonumber\\&&
-{{252\,251\,\pi^{2}}\over{9720}}+{679\,441\over93\,312}
\bigg] \delta_{\ell0} \nonumber\\
&&+ \bigg[
-{{100\,{\rm a_4}}\over{3}}+{{215\,\zeta(5)}\over{24}}
\nonumber\\&&
-{{83\,\pi^{2}\,\zeta(3)}\over{72}}-{{139\,\zeta(3)}\over{18}}
-{{25\,\ln ^{4}2}\over{18}}
\nonumber\\&&
+{{25\,\pi^{2}\,\ln ^{2}2}\over{18}}+{{298\,\pi^{2}\,\ln 2}\over{9}}
+{{239\,\pi^{4}}\over{2160}}
\nonumber\\&&
-{{17\,101\,\pi^{2}}\over{810}}-{28\,259\over5184}
\bigg] {m/m_r \over \kappa(2\ell+1)} (1 - \delta_{\ell0})
    \,,
\nonumber\\
\label{eq:c40}
\end{eqnarray}
where $a_4 = \sum_{n=1}^\infty 1/(2^n\,n^4) = 0.517\,479\,061\dots$. For the next-order
contribution $C_{50}$, there are only partial results up to now \cite{eides:04,eides:07}. Following
Ref.~\cite{mohr:16:codata}, we do not include partial results and estimate the uncertainty due to
absence of this term as $C_{50} = \pm 30\,\delta_{\ell0}$. The leading logarithmic contribution
$C_{62}$ was derived in Ref.~\cite{karshenboim:18} as
\begin{align}
C_{62} = -\frac23\,B_{40}\,,
\end{align}
where $B_{40}$ is the leading-order two-loop coefficient summarized in Table~\ref{tab:twoloop}.
Ref.~\cite{karshenboim:18} presented results also for the single-logarithmic contribution $C_{61}$
for the $nP$ states and the difference $C_{61}(nS)-C_{61}(1S)$.

\section{Nuclear recoil}
\label{sec:recoil}

The dominant part of the nuclear recoil effect is accounted for by $E_M$ in Eq.~(\ref{eq3}) and by
the reduced-mass prefactors in previous formulas. Beyond that, there are a number of further recoil
corrections. The first one is the nuclear recoil correction of order $(\Za)^{\ge5}$ and of first
order in $m/M$,
\begin{align}\label{rec:1}
E_{\rm REC} = &\  \frac{m}{M}\, \frac{(\Za)^5}{\pi\, n^3}\,
 \biggl[ \left(\frac{m_r}{m}\right)^3 \ln (\Za)^{-2}\,D_{51}
  \nonumber \\ &
 + \left(\frac{m_r}{m}\right)^3 \,D_{50} + (\Za)\, D_{60}+ (\Za)^2\,G_{\rm REC}(\Za)\biggr]\,,
\end{align}
where $G_{\rm REC}(\Za)$ is the higher-order remainder containing all higher orders in $\Za$.
Coefficients of the $Z\alpha$ expansion in Eq.~(\ref{rec:1}) are reviewed in Ref.~\cite{eides:01}
and summarized in Table~\ref{tab:rec}. The higher-order remainder $G_{\rm REC}$ has an expansion of
the form
\begin{align}\label{rec:1b}
G_{\rm REC}(\Za) = D_{72}\,\ln^2(\Za)^{-2} + D_{71}\,\ln^2(\Za) + D_{70} + \ldots\,,
\end{align}
where $D_{72} = -11/60\, \delta_{\ell0}$ \cite{pachucki:99:prab,melnikov:99} and the next two
coefficients were obtained by fitting numerical results in
Refs.~\cite{yerokhin:15:recprl,yerokhin:16:recoil}
\begin{align}
&\ D_{71}(1S)  = 2.919\,(10)\,,\ \ \
D_{70}(1S)  = -1.32\,(10)\,,\\
&\ D_{71}(2S)  = 3.335\,(10)\,,\ \ \
D_{70}(2S)  = -0.26\,(6)\,,\\
&\  D_{71}(2P_{1/2}) = 0.149\,(5)\,,\ \ \
D_{70}(2P_{1/2})  = -0.035\,(15)\,.
\end{align}
Numerical, all-order in $\Za$ results for the higher-order remainder $G_{\rm REC}$ are obtained in
Refs.~\cite{yerokhin:15:recprl,yerokhin:16:recoil} and summarized in Table~\ref{tab:rec:ho}. In the
present review we do not include results for the finite nuclear size correction to $E_{\rm REC}$
obtained in Refs.~\cite{yerokhin:15:recprl,yerokhin:16:recoil}, since this effect is partly
included in calculations of nuclear polarizability summarized in the next section.

The relativistic recoil corrections of second order in the mass ratio is
\cite{erikson:77,pachucki:95:jpb,sapirstein:90:kin},
\begin{align}\label{rec:3}
E_{\rm REC,2} = &\  \left(\frac{m}{M}\right)^2\, \frac{(\Za)^4}{n^3}\,
 \biggl[ \frac{3}{4n} - \frac1{2l+ 1} + \frac12\,\delta_{\ell0}\,\delta_{I,1/2}
 \nonumber \\ &
  - (\Za)\,\frac2{\pi}\,\left( 1+ \frac{m}{M}\ln \frac{m}{M}\right)\,\delta_{\ell0}
 \biggr]\,.
\end{align}
The first part of this correction $\propto\!(\Za)^4$ depends on the nuclear spin $I$, which is the
consequence of the choice of the definition of the point-like particle with a spin $I$. For $I > 1$
such a definition is not commonly established, so we ascribe an uncertainty of $\pm
\frac12\,\delta_{\ell0}$ relative to the square brackets in the above formula. This part agrees
with the $(\Za)^4 (m/M)^2$ term contained in Eq.~(25) of the CODATA review \cite{mohr:16:codata}.
The second part of this correction $\propto\!(\Za)^5$ is the Erickson formula (see the last line of
Eq. (27) in Ref.~\cite{mohr:16:codata}) expanded in $m/M$. This formula is derived for the
spin-$1/2$ nucleus; its dependence on nuclear spin is not known. However, we assume the
corresponding uncertainty to be negligible.

An additional recoil contribution arises for the $P$ (and higher-$l$) states because of mixing of
the fine-structure sublevels by the hyperfine-structure (hfs) interaction. This contribution is
also known as the off-diagonal hfs shift. It depends on the nuclear spin $I$ and the nuclear
magnetic moment $\mu$ and is given, for the $nP$ states \cite{brodsky:67,horbatsch:16}, by
\begin{align}\label{rec:4}
E_{\rm REC,hfs}(nP) = &\  \left(\frac{m}{m_p}\right)^2\, \frac{\alpha^2 (\Za)^2}{n^3}\,
 \nonumber \\ & \times
 \Big( \frac{\mu}{\mu_N} \Big)^2 \frac{2I(I+1)}{81}\,(-1)^{j+1/2}\,\delta_{\ell1}\,,
\end{align}
where $\mu_N = |e|/(2\,m_p)$ is the nuclear magneton and $m_p$ is the proton mass. This correction
shifts the $2P_{1/2}$ centroid energy by $-1.88$~kHz for hydrogen, by $-0.47$~kHz for deuterium,
and by $-4.36$~kHz for $^3$He. We note that this correction was not included in the definition of
the energy levels in the CODATA review \cite{mohr:16:codata} and needed to be accounted for
together with the hyperfine structure. Corrections to Eq.~(\ref{rec:4}) are assumed to be
suppressed by $\alpha/\pi$, which is included into uncertainty.

Furthermore, there is the radiative recoil correction
\cite{pachucki:95,czarnecki:01,eides:01:pra,pachucki:99:prab}
\begin{align}\label{rec:2}
E_{\rm RREC} = &\ \frac{m}{M}\, \left(\frac{m_r}{m}\right)^3\, \frac{\alpha(\Za)^5}{\pi^2 n^3}\,\delta_{\ell0}
\biggl[ 6\,\zeta(3)-2\pi^2\ln 2
  \nonumber \\ &
+ \frac{35\pi^2}{36} - \frac{448}{27} + \frac{2}{3}\,\pi(\Za)\ln^2(\Za)^{-2} \biggr]\,.
\end{align}
Following Ref.~\cite{mohr:12:codata}, we ascribe to this correction an uncertainty of
$10(\Za)\ln(\Za)^{-2}$ relative to the square brackets in the above equation.

\begin{table*}
\caption{Coefficients of the $Z\alpha$ expansion of the nuclear recoil correction in Eq.~(\ref{rec:1}). \label{tab:rec}}
\begin{ruledtabular}
\begin{tabular}{lccccc}
\multicolumn{1}{c}{Term} &  \multicolumn{1}{c}{} & \multicolumn{1}{c}{$1S$} & \multicolumn{1}{c}{$2S$}& \multicolumn{1}{c}{$2P_{1/2}$}\\
\hline\\[-5pt]
$D_{51}$ & $\frac{1}{3}  \,\delta_{\ell0}$ &  $\frac{1}{3}  $  & $\frac{1}{3}  $ & 0 \\[5pt]
$D_{50}$ & $-\frac{8}{3}\,\ln k_0 + \frac{14}{3} \left[ 1 -\frac1{42} -\frac1{2n} + \ln \frac{2}{n} + \psi(n+1)-\psi(1)\right] \delta_{\ell0} $
             & 2.165\,899\,582  & 2.890\,835\,841 & $-$0.308\,844\,332 \\[5pt]
         & $-\frac73\, [l(l+1)(2l+1)]^{-1}(1-\delta_{\ell0})$ &       \\[5pt]
$D_{60}$ & $\left( 4\ln 2-\frac{7}{2}\right)\pi\,\delta_{\ell0}+ 2\pi \left[ 3- \frac{l(l+1)}{n^2}\right]\, [(4l^2-1)(2l+3)]^{-1} (1-\delta_{\ell0})$
             & $-$2.285\,229\,926  & $-$2.285\,229\,926   & 1.047\,197\,551 \\
\end{tabular}
\end{ruledtabular}
\end{table*}

\begin{table}
\caption{Numerical results for the recoil higher-order remainder function in Eq.~(\ref{rec:1}).
\label{tab:rec:ho}}
\begin{ruledtabular}
\begin{tabular}{lddd}
\multicolumn{1}{c}{$Z$} & \multicolumn{1}{c}{$1S$} & \multicolumn{1}{c}{$2S$}& \multicolumn{1}{c}{$2P_{1/2}$}\\
\hline\\[-5pt]
  1    & 9.720\,(3) &   14.899\,(3) & 1.509\,7\,(2) \\
  2    & 10.390\,(1) &  15.010\,(1) & 1.307\,39\,(5)\\
  3    & 10.4803\,(9) & 14.7806\,(9)& 1.192\,04\,(2)\\
  4    & 10.4155\,(6) & 14.4926\,(6)& 1.112\,68\,(2)\\
  5    & 10.2944\,(4) & 14.2013\,(4)& 1.053\,21\,(2)\\
\end{tabular}
\end{ruledtabular}
\end{table}

\section{Nuclear size and polarizability}

It is customary in the literature to consider separately the finite nuclear size (fns) effect (also
known as the elastic part of the nuclear structure) and the nuclear polarizability (also known as
the inelastic nuclear structure).  To a large extent, the separate treatment is due to the fact
that the fns correction can be obtained numerically from the Dirac equation, whereas calculations
of the nuclear polarizability are much more complicated. However, it was shown \cite{friar:97:b,
pachucki:11:mud,pachucki:18} that for light atoms, there is significant cancelation between the fns
effects and the polarizability corrections. Moreover, it turned out that some of the nuclear model
dependence of the individual corrections cancels out in the sum. Because of this, it is desirable
to keep these contributions together and address them on the same footing. We thus consider the sum
of the fns correction $E_{\rm fns}$ and the polarizability correction $E_{\rm pol}$,
\begin{align}
E_{\rm nucl} = E_{\rm fns} + E_{\rm pol} = \sum_{i \ge 4} E_{\rm nucl}^{(i)}\,,
\end{align}
where the upper index $i$ indicates the order in $Z\alpha$.

\subsection{ $\bm{(\Za)^4}$ nuclear contribution}

The leading-order nuclear contribution comes solely from the finite nuclear size. It is given for
an arbitrary hydrogen-like system by a simple formula,
\begin{align} \label{eq:fns:lo}
E^{(4)}_{\rm nucl} = E^{(4)}_{\rm fns} =
 \frac23 \frac{(\Za)^4}{n^3}
\left(\frac{m_r}{m}\right)^3
 R_C^2\,\delta_{\ell0}\,,
\end{align}
where $R_C$ is the root-mean-square (rms) charge radius of the nucleus
\begin{align}
  R_C^2 = \int d^3 r\; r^2\,\rho(r)\,,
\end{align}
and $\rho(r)$ is the nuclear charge distribution.

The higher-order nuclear contributions are specific for each nucleus. We start our consideration
with hydrogen, which is a special case since proton is the only non-composite (one-nucleon)
nucleus.

\subsection{ $\bm{(\Za)^5}$ nuclear contribution for hydrogen}

If we assume that the nucleus has a fixed charge density distribution, then the $(\Za)^5$ nuclear
correction is given by the two-Coulomb exchange amplitude. The resulting fns correction is
\cite{friar:79:ap}
\begin{align}
  E^{(5)}_{\rm fns} = -\frac{1}{3}\,\frac{(\Za)^5}{n^3} \left(\frac{m_r}{m}\right)^3\,R_Z^3
  \,\delta_{\ell0}\,, \label{zemach}
\end{align}
where $R_Z$ is the third Zemach moment
\begin{equation} \label{rZ}
R_Z^3 =  \int d^3r_1\int d^3r_2\,\rho(r_1)\,\rho(r_2)\,|\vec r_1-\vec r_2|^3\,.
\end{equation}
The numerical value for the proton is $R_{pZ} \equiv R_Z({\rm H}) = 1.41(2)$~fm, which is the
average of two results derived from the electron-positron scattering \cite{friar:04,distler:11}.

A more detailed consideration shows, however, that a nucleus cannot generally be treated as a rigid
body, because it is polarized by the surrounding electron. This gives rise to the so-called nuclear
polarizability contribution. The proton polarizability correction is usually calculated as the
forward two-photon exchange amplitude, expressed via dispersion relations in terms of the inelastic
scattering amplitude, which in turn is accessible in experiments.

The recent evaluation of the proton $(\Za)^5$ nuclear contribution \cite{tomalak:18} yields the
result of $-0.1092\,(120)$~kHz for the hydrogen $1S$ state, which agrees with the previous (elastic
$+$ polarizability) value adopted by CODATA \cite{mohr:16:codata} of $-0.10(1)$~kHz. The result
\cite{tomalak:18} can be conveniently parameterized in terms of the effective proton radius
$R_{pF}$, which is introduced in analogy with Eq.~(\ref{zemach}),
\begin{align} \label{pFa}
  E^{(5)}_{\rm nucl}({\rm H}) = -\frac{1}{3}\,\frac{(\Za)^5}{n^3}\, \left(\frac{m_r}{m}\right)^3
  \,R_{pF}^3\,\delta_{\ell0}\,,
\end{align}
with
\begin{equation} \label{pF}
R_{pF} = 1.947\,(75)\;{\rm fm}\,.
\end{equation}
We note that for the proton there is no cancelation between the elastic and polarizability
contributions, in contrast to the composite nuclei.

\subsection{ $\bm{(\Za)^5}$ nuclear contribution for composite nuclei}

For compound nuclei consisting of several nucleons, the Zemach fns correction (\ref{zemach})
cancels out in a sum with the corresponding nuclear structure contribution \cite{pachucki:18}.
However, it survives in the contribution induced by the interaction with individual nucleons. In
the result, we write the total nuclear structure correction $E^{(5)}_{\rm nucl}$ (known also as the
two-photon exchange correction) for a composite nuclei as
\begin{align}
  E^{(5)}_{\rm nucl} = E^{(5)}_{\rm pol} -\frac{1}{3}\,\frac{\alpha^2(\Za)^3}{n^3}\,
  \big[Z\,R_{pF}^3 + (A-Z)\,R_{nF}^3\big]\,\delta_{\ell0}\,,
\end{align}
where the first term $E^{(5)}_{\rm pol}$ is the intrinsic nuclear polarizability and the second
term is the contribution of individual nucleons. In the above equation, $R_{pF}$ is the effective
proton radius given in Eq.~(\ref{pF}), $R_{nF}$ is an analogous effective radius for the neutron,
and $A$ is the mass number. We extract $R_{nF}$ from the calculation of Tomalak (Table II of
Ref.~\cite{tomalak:18}), with the result
\begin{equation} \label{nF}
R_{nF} = 1.43\,(16)\;{\rm fm}\,.
\end{equation}

The nuclear polarizability correction $E^{(5)}_{\rm pol}$ is dominated by the electric dipole
excitations and is given by \cite{pachucki:11:mud,hernandez:14,pachucki:18}
\begin{align}
  E^{(5)}_{\rm pol} =&\  -\alpha^2\,\phi^2(0)\,\frac{2}{3}\,
  \biggl\langle\phi_N\biggl|\,\vec d\,\frac{1}{H_N-E_N}
  \biggl[\frac{19}{6}
   \nonumber \\ &\
  + 5\,\ln\frac{2\,(H_N-E_N)}{m} \biggr]\,\vec d\,\biggr|\phi_N\biggr\rangle\,
  \nonumber \\ &\
  -\frac{\pi}{3}\,\alpha^2\,\phi^2(0)\,\sum_{i,j=1}^Z\langle\phi_N||\vec R_i-\vec R_j|^3|\phi_N\rangle
  \nonumber \\&\
  + {\mbox{\rm many small corrections}}\,, \label{epol}
\end{align}
where $\vec{d}$ is the electric dipole operator divided by the elementary charge, $H_N$ and $E_N$
are the nuclear Hamiltonian and its eigenvalue, $\phi_N$ and $\phi$ are the nuclear and electronic
wave functions, and $\vec R_i$ is the position vector of $i$th proton in the nucleus. The second
term in Eq.~(\ref{epol}) is the remainder of the Zemach fns correction (\ref{zemach}) for a
composite nuclei.

For atoms with $Z \le 5$, the nuclear polarizability correction has been investigated only for
deuterium, helium, and some neutron-rich isotopes of Li and Be. For deuteron, the two-photon
nuclear polarizability was calculated in Ref.~\cite{friar:97:b} and recently reanalysed in
Ref.~\cite{pachucki:18},
\begin{equation}
  E^{(5)}_{\rm pol}({\rm D}) = -21.78\,\,\frac{\delta_{\ell0}}{n^3}~h\,{\rm kHz} \pm 1\%\,.
\end{equation}
For helium, the nuclear polarizability correction was calculated in
Ref.~\cite{pachucki:07:heliumnp}, with the result
\begin{align}
  E^{(5)}_{\rm pol}(^4{\rm He}) &\ = -32.1\,\,\frac{\delta_{\ell0}}{n^3}~h\,{\rm kHz} \pm 10\%\,,\\
  E^{(5)}_{\rm pol}(^3{\rm He}) &\ = -55.2\,\,\frac{\delta_{\ell0}}{n^3}~h\,{\rm kHz} \pm 10\%\,.
\end{align}

For stable isotopes with $Z = 3$, 4, and 5, we use the following estimate
\begin{align}
E^{(5)}_{\rm pol} \approx - \frac{E_{\rm fns}}{1000} \pm 100\%\,,
\end{align}
which was obtained in Ref.~\cite{yerokhin:15:Hlike} basing on an analysis of available results
throughout the whole $Z$ sequence.

\subsection{ $\bm{(\Za)^6}$ nuclear contribution}

The $(\Za)^6$ nuclear contribution arises from the three-photon exchange between electron and the
nucleus. The corresponding fns correction is known in the nonrecoil limit and is given for the $nS$
and $nP_{1/2}$ ($\kappa=1$) states by \cite{friar:79:ap,pachucki:18}
\begin{align} \label{fns6}
  E^{(6)}_{\rm fns} =&\  \frac{(\Za)^6}{n^3}\,R_C^2
%  \left(\frac{m_r}{m}\right)^3
  \Bigg\{
-\frac{2}{3} \,
 \biggl[\frac{9}{4n^2}
 - 3 -\frac{1}{n}
 \nonumber \\ &
   +2\,\gamma_E -\ln\frac{n}{2}+\Psi(n)
        +\ln\big(m R_{C2}\,Z\,\alpha\big)\biggr] \delta_{\ell0}
 \nonumber \\ &
 + \frac{1}{6}\,\biggl(1-\frac{1}{n^2}\biggr)\, \delta_{\kappa1}
 \Bigg\}\,,
\end{align}
where $R_{C2}$  is the effective nuclear charge radii that encodes the high-momentum contribution
(for exact definition see Ref.~\cite{pachucki:18}). The effective nuclear radii $R_{C2}$ has the
numerical value close to $R_C$ and depends on the model of the nuclear charge distribution. We use
the result obtained in Ref.~\cite{pachucki:18} for the exponential model,
\begin{align}
R_{C2}/R_C = 1.068\,497\,,
\end{align}
which does not depend on nuclear charge. It was shown in Ref.~\cite{pachucki:18} that the
dependence on $R_{C2}$ in Eq.~(\ref{fns6}) cancels out in the sum with the corresponding nuclear
polarizability correction, so the model dependence of $R_{C2}$ does not contribute to the
uncertainty.

The $(\Za)^6$ nuclear polarizability is practically unknown for the electronic atoms. The only
available results are estimates from Ref.~\cite{pachucki:18} for hydrogen
\begin{equation}
  E^{(6)}_{\rm pol}({\rm H}) = 0.393\,\frac{\delta_{\ell0}}{n^3}~h\,{\rm kHz} \pm 100\%\,,
\end{equation}
and deuterium
\begin{equation}
  E^{(6)}_{\rm pol}({\rm D}) = -0.541\,\frac{\delta_{\ell0}}{n^3}~h\,{\rm kHz} \pm 75\%\,.
\end{equation}

It is remarkable that for hydrogen, the three-photon nuclear polarizability dominates over the
two-photon polarizability. The reason for this is that $E^{(6)}_{\rm pol} \propto (\Za)^6\,R_C^2$
whereas $E^{(5)}_{\rm nucl}({\rm H}) \propto (\Za)^5\,R_C^3$, so that the two-photon exchange is
effectively suppressed by a parameter $mR_C/(\Za) \ll 1$. For all atoms other than hydrogen, the
two-photon exchange is dominated by the electric dipole polarizability $\propto (\Za)^5\,R_C^2$
and, therefore, the three-photon polarizability is smaller than the two-photon one, as usually
expected. We estimate the uncertainty due to the unknown three-photon nuclear polarizability for
nuclei with $Z = 2-5$ to be 10\% of the corresponding two-photon polarizability.

\subsection{Radiative fns correction}

The leading radiative fns correction is of order $\alpha(\Za)^5$ and nonzero only for $S$ states
(see review \cite{eides:01} for details),
\begin{align} \label{eq:fns:5}
  E^{(5)}_{\rm fns, rad} =&\, \frac23 \frac{\alpha(\Za)^5}{n^3} \left(\frac{m_r}{m}\right)^3
  R_C^2\,\big(4\ln 2-5\big)\,\delta_{\ell0}\,.
\end{align}

The next-order radiative fns correction for the $S$ states is known only partially
\cite{milstein:03:fns,jentschura:03:jpa,yerokhin:11:fns},
\begin{align} \label{eq:fns:6}
  E^{(6)}_{\rm fns, rad}(nS) =&\, \frac23 \frac{\alpha\,(\Za)^6}{\pi\,n^3}\,
  R_C^2\,\Big[ -\frac23\,\ln^2 (\Za)^{-2}
% \nonumber \\ &
    + \ln^2 (mR_{C})
    \Big]\,.
\end{align}
In the above formula we keep only the squared logarithms and do not include some higher-order terms
derived in Ref.~\cite{milstein:03:fns}, because the term $\propto \ln (\Za)^{-2}$ is not known and
expected to be of similar magnitude as the omitted terms. The result for the $P$ states
\cite{milstein:03:fns,jentschura:03:jpa,yerokhin:11:fns} is
\begin{align} \label{eq:fns:7}
  E^{(6)}_{\rm fns, rad}&\,(nP_{1/2}) =
   \frac16 \frac{\alpha\,(\Za)^6}{\pi\,n^3}\,R_C^2\,
\left(1-\frac1{n^2}\right)
 \nonumber \\ & \times
    \Bigg[ \frac89\,\ln (\Za)^{-2}  -\frac89 \ln 2 + \frac{166}{135}
% \nonumber \\ &
 + \frac{4n^2}{n^2-1}{\cal N}(nP) \Bigg]\,.
\end{align}
The uncertainty of Eqs.~(\ref{eq:fns:6}) and (\ref{eq:fns:7}) was evaluated by comparing with
results of the more complete treatment \cite{yerokhin:11:fns}.

\subsection{Nuclear self-energy}

The nuclear self-energy correction was derived in Ref.~\cite{pachucki:95:pra}, with the result
\begin{align}\label{nse}
E_{\rm NSE} = &\ \left(\frac{m}{M}\right)^2\, \frac{4Z(\Za)^5}{3\pi n^3}\,
  \nonumber \\ &
\times \biggl[ \ln\left(\frac{M}{m(\Za)^2}\right)\,\delta_{\ell0}
- \ln k_0(n,l)
\biggr]\,.
\end{align}
It should be noted that there is some ambiguity associated with this correction since the nuclear
self-energy contributes not only to the Lamb shift but to the nuclear charge radius and the nuclear
magnetic moment. Specifically, addition of an arbitrary constant in the brackets of Eq.~(\ref{nse})
is equivalent to changing the definition of the nuclear charge radius. This implies that the
presently used definition of the nuclear charge radius (through the slope of the Sachs form-factor)
is ambiguous on the level of a constant in the brackets of Eq.~(\ref{nse}). This issue was pointed
out in Ref.~\cite{pachucki:95:pra} (together with the suggestion for a rigorous definition of the
nuclear charge radius) but did not attracted attention of the community up to now. In order to
quantify this ambiguity, we ascribe to $E_{\rm NSE}$  an uncertainty of 0.5 in the square brackets,
as in Ref.~\cite{mohr:12:codata}. The numerical value of this uncertainty is $0.2$~kHz for the
hydrogen $1S$ state, which can be disregarded at present but might become relevant in the future.

\section{Numerical results}

In order to obtain numerical results for the Lamb shift and the transition energies, we need to
specify values of fundamental constants and nuclear parameters. In the present review we use the
charge radii of the proton and the deuteron as derived from the muonic atoms
\cite{antognini:13,pohl:16} ($R_p = 0.84087\,(39)\ \mbox{\rm fm}$ and $R_d = 2.12562\,(78)\
\mbox{\rm fm}$) and the corresponding value of the Rydberg constant from Ref.~\cite{horbatsch:16},
\begin{align}\label{rydberg}
 c\,R_{\infty} = 3\,289\,841\,960\,248.9\,(3.0)\ \mbox{\rm kHz}\,.
\end{align}
It should be mentioned that the exact values of $R_p$, $R_d$, and $R_{\infty}$ are under debates at
present. In particular, the Rydberg constant of Eq.~(\ref{rydberg}) differs from the value
recommended by CODATA 2014 \cite{mohr:16:codata} by $5.5\,\sigma$. On the level of the present
experimental accuracy, this controversy is relevant only for hydrogen and deuterium and can be
disregarded for heavier atoms.

The nuclear charge radii for elements with $Z>1$ are taken as follows. For $^3$He and $^4$He, we
use values by Sick \cite{sick:08,sick:15}; for $^6$Li and $^7$Li isotopes, values from
Ref.~\cite{nortenhauser:11}; for other atoms, values from Ref.~\cite{angeli:13}. The nuclear masses
are taken for hydrogen from Ref.~\cite{heisse:17}, for deuterium and helium isotopes from
Ref.~\cite{mohr:16:codata}, and for all other nuclei from Ref.~\cite{wang:12}. Nuclear magnetic
moments are taken from Ref.~\cite{stone:05}. The fine-structure constant is \cite{mohr:16:codata}
\begin{align}
\alpha = 1/137.035\,999\,139\,(31)\,.
\end{align}

The individual contributions to the Lamb shift for two experimentally most interesting cases,  H
and He$^+$, are listed in Table~\ref{tab:EL}. The results for the QED and the leading fns
correction are presented in the nonrecoil limit (i.e., with $m_r\to 1$). The contribution due to
the reduced mass in all formulas is summed up and tabulated separately as the relativistic reduced
mass (RRM) correction. The uncertainty of the fns correction is due to the uncertainty of the
nuclear charge radius $R_C$, whose values are specified in the table. The total results for the
Lamb shift $E_L$ are given with two uncertainties. The first one is the theoretical uncertainty,
whereas the second one comes from the uncertainty of the nuclear charge radius.

We observe that for the hydrogen Lamb shift, the theoretical uncertainty is twice larger than the
uncertainty due to the proton charge radius (as extracted from muonic hydrogen). The two largest
theoretical uncertainties come from (i) the two-loop self-energy and (ii) the three-loop QED
correction. As compared to the previous CODATA review \cite{mohr:16:codata}, the main change is due
to our reanalysis of the two-loop QED effects; it shifted the theoretical value by one half of the
previous uncertainty and improved the accuracy by a factor of 1.5.

For helium, the uncertainty of the Lamb shift is presently dominated by the uncertainty from the
nuclear radius. But this is likely to change once the results of the muonic helium experiment are
evaluated \cite{pohl:talk,schmidt:18}.

Table~\ref{tab:energy} presents theoretical results for the $2S$--$1S$ and $2S$--$2P_{1/2}$
transition energies in hydrogen and light hydrogen-like ions. Theoretical predictions are given
with two uncertainties. The first one is the theoretical uncertainty, whereas the second one is
induced by uncertainties of nuclear radii and masses. The uncertainty due to the Rydberg constant
$R_{\infty}$ is not included. Theoretical predictions are compared with available experimental
results for the $2S$--$2P_{1/2}$ Lamb shift in hydrogen, helium and lithium. We do not present a
comparison with the hydrogen $1S$--$2S$ experimental results \cite{parthey:10,matveev:13} since the
value of the Rydberg constant (\ref{rydberg}) is derived from the comparison of theory and these
experiments. For the same reason we do not include the uncertainty due to Rydberg constant in the
theoretical predictions.

In summary, theoretical calculations of the Lamb shift in hydrogen and light hydrogen-like ions are
required for the determination of the Rydberg constant. In the present work we summarized the
present status and recent developments of theoretical calculations of QED and nuclear effects,
critically evaluating uncertainties of all contributions.

%%%%%%%%%%%%%%%%%%%%%%%%%%%%%%%%%%%%%%%%%%%%%%%%%%%%%%%%%%%%%%%%%%%%%%%%%%%%%%%%%

\begin{acknowledgments}
The authors are grateful to A.~Kramida for pointing out a mistake in an early version of the
manuscript. V.A.Y. acknowledges support by the Ministry of Education and Science
  of the Russian Federation Grant No. 3.5397.2017/6.7.
K.P. acknowledges support by the National Science Center (Poland) Grant No. 2017/27/B/ST2/02459.
 V.P. acknowledges support from the Czech Science Foundation - GA\v{C}R (Grant No. P209/18-00918S).
\end{acknowledgments}

\begin{table*}
\caption{Individual contributions to the Lamb shift $E_L$, in MHz. Abbreviations are as follows:
``SE" is the one-loop self-energy, ``Ue'' is the Uehling one-loop vacuum polarization,
``WK" is the Wichmann-Kroll one-loop vacuum-polarization,
``Ue($\mu$had)'' is the Uehling muon and hadronic vacuum polarization,
``SESE" is the two-loop self-energy,
``SEVP" is the electron self-energy with vacuum-polarization insertions, ``VPVP" is the two-loop vacuum-polarization,
``QED(ho)" is the three-loop QED correction,
``RRM'' is the relativistic reduced mass correction (see text), ``REC'' is the recoil correction $E_{\rm REC}$,
``REC(ho)'' is the sum of higher-order recoil corrections $E_{\rm REC,2}$, $E_{\rm REC,hfs}$, and $E_{\rm RREC}$,
``FNS'' is the leading-order fns correction $E_{\rm nucl}^{(4)}$,
``NUCL5'' is the $(\Za)^5$ nuclear correction $E_{\rm nucl}^{(5)}$,
``NUCL6'' is the $(\Za)^6$ nuclear correction $E_{\rm nucl}^{(6)}$,
``FNS(rad)'' is the radiative fns correction $E_{\rm fns,rad}$,
``NSE'' is the nuclear self-energy correction $E_{\rm NSE}$.
\label{tab:EL}}
\begin{ruledtabular}
\begin{tabular}{lddd}
  & \multicolumn{1}{c}{$1S$}  & \multicolumn{1}{c}{$2S$}  & \multicolumn{1}{c}{$2P_{1/2}$} \\
\hline\\[-5pt]
\multicolumn{3}{l}{
  $Z = 1,\ \mbox{\rm $^{1}$H },\  R_C = 0.840\,87\,(39)\ \mbox{\rm fm}, \  M/m = 1\,836.152\,673\,346\,(81)$}
 \\[2pt]
SE                   & 8\,396.453\,556\,(1)  & 1\,072.958\,455   & -12.858\,661\,(1) \\
Ue                   & -215.170\,186       & -26.897\,303        & -0.000\,347 \\
WK                   & 0.002\,415          & 0.000\,302          & 0 \\
Ue($\mu$had)         & -0.008\,48\,(8)     & -0.001\,06\,(1)     & 0 \\
SESE                 & 2.335\,0\,(13)      & 0.292\,48\,(16)     & 0.027\,253\,(4) \\
SEVP                 & 0.288\,39\,(16)     & 0.036\,015\,(20)    & -0.001\,241 \\
VPVP                 & -1.895\,224         & -0.236\,911         & -0.000\,003 \\
QED(ho)              &  0.001\,83\,(96)    &  0.000\,23\,(12)    & -0.000\,216 \\
RRM                  & -12.765\,917        & -1.633\,931         & 0.011\,741 \\
REC                  & 2.402\,830          & 0.340\,469          & -0.016\,656 \\
REC(ho)              & 0.013\,16\,(74)     & -0.003\,227\,(92)   & -0.001\,335\,(4) \\
FNS                  & 1.107\,6\,(10)      & 0.138\,45\,(13)     & 0 \\
NUCL5                & -0.000\,109\,(1)    & -0.000\,014         & 0 \\
NUCL6                & 0.001\,07\,(39)     & 0.000\,140\,(49)    & 0.000\,001 \\
FNS(rad)             & -0.000\,135\,(1)    & -0.000\,017         & 0 \\
NSE                  & 0.004\,63\,(16)     & 0.000\,585\,(20)    & 0.000\,001\,(20) \\
Total                & 8\,172.770\,4\,(18)(10) & 1\,044.994\,66\,(23)(13) & -12.839\,463\,(21)(0) \\
\\[-5pt]
\multicolumn{3}{l}{
  $Z = 2,\ \mbox{\rm $^{4}$He$^+$},\  R_C = 1.6810\,(40)\ \mbox{\rm fm}, \  M/m = 7\,294.299\,541\,36\,(24)$}
 \\[2pt]
SE                   & 111\,054.170\,69\,(1) & 14\,257.035\,60\,(2)& -204.794\,17\,(2) \\
Ue                   & -3\,415.099\,45       & -426.952\,77        & -0.022\,109 \\
WK                   & 0.152\,06             & 0.019\,01           & 0.000\,002 \\
Ue($\mu$had)         & -0.135\,7\,(12)       & -0.016\,97\,(15)    & 0 \\
SESE                 & 32.569\,(64)          & 4.095\,9\,(80)      & 0.440\,50\,(25) \\
SEVP                 & 4.956\,8\,(88)        & 0.617\,9\,(11)      & -0.020\,086 \\
VPVP                 & -30.047\,28           & -3.756\,393\,(1)    & -0.000\,210 \\
QED(ho)              &  0.029\,(31)          &  0.003\,6\,(38)     & -0.003\,468 \\
RRM                  & -41.915\,19           & -5.393\,65          & 0.046\,950 \\
REC                  & 17.676\,28            & 2.533\,38           & -0.130\,835 \\
REC(ho)              & -0.121\,(10)          & -0.020\,0\,(13)     & 0.000\,549 \\
FNS                  & 70.82\,(34)           & 8.853\,(42)         & 0 \\
NUCL5                & -0.034\,6\,(32)       & -0.004\,33\,(40)    & 0 \\
NUCL6                & 0.152\,0\,(35)        & 0.020\,66\,(43)     & 0.000\,354 \\
FNS(rad)             & -0.017\,43\,(39)      & -0.002\,179\,(50)   & 0.000\,007 \\
NSE                  & 0.018\,75\,(65)       & 0.002\,372\,(82)    & 0.000\,005\,(82) \\
Total                & 107\,693.18\,(7)(34)  & 13\,837.035\,(9)(42) & -204.482\,51\,(26)(0) \\
\\[-5pt]
\end{tabular}
\end{ruledtabular}
\end{table*}

\begin{table*}
\caption{Theoretical transition energies of light hydrogen-like atoms (in GHz), in comparison with available experimental results.
\label{tab:energy}}
\begin{ruledtabular}
\begin{tabular}{lllcdd}
  $Z$  &  \multicolumn{1}{c}{}  &  \multicolumn{1}{c}{$R_C\,$[fm]}  &  \multicolumn{1}{c}{$M/m$}
        &  \multicolumn{1}{c}{$2S$--$1S$}  &  \multicolumn{1}{c}{$2S$--$2P_{1/2}$} \\
\hline\\[-5pt]
  1 & \mbox{\rm $^{  1}$H } & 0.84087\,(39) & 1\,836.152\,673\,346\,(81) & 2\,466\,061.413\,1869\,(18)(10) & 1.057\,834\,12\,(23)(13) \\
    &                       &               &                      &                                       & 1.057\,847\,(9)^a\\
  1 & \mbox{\rm $^{  2}$D } & 2.12562\,(78) & 3\,670.482\,967\,85\,(13) & 2\,466\,732.407\,5345\,(17)(52)  & 1.059\,219\,91\,(21)(65) \\
  2 & \mbox{\rm $^{  4}$He$^+$} & 1.6810\,(40) & 7\,294.299\,541\,36\,(24) & 9\,868\,561.006\,31\,(7)(34)      & 14.041\,517\,(9)(42) \\
    &                       &              &                     &                                         & 14.041\,13\,(17)^b \\
  2 & \mbox{\rm $^{  3}$He$^+$} & 1.973\,(16) & 5\,495.885\,279\,22\,(27) & 9\,868\,118.3826\,(1)(16) & 14.043\,96\,(1)(20) \\
  3 & \mbox{\rm $^{  6}$Li$^{2+}$} & 2.589\,(39) & 10\,961.898\,642\,0\,(83) & 22\,206\,430.550\,(1)(26) & 62.734\,2\,(1)(32) \\
    &                       &             &                     &                       & 62.765\,(21)^c \\
  3 & \mbox{\rm $^{  7}$Li$^{2+}$} & 2.444\,(42) & 12\,786.392\,271\,(11) & 22\,206\,719.625\,(1)(26) & 62.723\,1\,(1)(33) \\
  4 & \mbox{\rm $^{  9}$Be$^{3+}$} & 2.519\,(12) & 16\,424.205\,51\,(16) & 39\,482\,224.239\,(4)(24) & 179.771\,9\,(5)(30) \\
  5 & \mbox{\rm $^{ 11}$B$^{4+}$}  & 2.406\,(29) & 20\,063.737\,33\,(78) & 61\,697\,635.70\,(1)(14) & 404.523\,(1)(17) \\
\end{tabular}
\end{ruledtabular}
$^a$ Experiment by Lundeen and Pipkin \cite{lundeen:86}; the original result is shifted by 1.88~kHz due to the off-diagonal hfs correction,
in order to comply with the present definition of the Lamb shift,\\
$^b$ Experiment by van Wijngaarden et al.~\cite{wijngaarden:00},\\
$^c$ Experiment by Leventhal \cite{leventhal:75}.\\
\end{table*}

%%%%%%%%%%%%%%%%%%%%%%%%%%
%\bibliography{/home/krp/hydrogen/yerokhin_review//hfst}
%%%%%%%%%%%%%%%%%%%%%%%%%%
%\bibliographystyle{c:/-a-/papers/bibtex/my}
%\bibliography{c:/-a-/papers/bibtex/hfst}

\end{document}